\documentclass[conference]{IEEEtran}
\IEEEoverridecommandlockouts
\ifCLASSINFOpdf
\else
\fi

\usepackage{marvosym}
\usepackage{cite}
\usepackage{amsmath,amssymb,amsfonts}
\usepackage{graphicx}
\usepackage{textcomp}
\usepackage{xcolor}

\usepackage{booktabs}
\usepackage{multirow}

\begin{document}
%
\title{Orthogonal Hyper-category Guided  Multi-interest Elicitation  for Micro-video Matching}

\author{\IEEEauthorblockN{Beibei Li}
\IEEEauthorblockA{\textit{College of Computer Science} \\
\textit{Chongqing University}\\
Chongqing, China \\
libeibeics@cqu.edu.cn}
\and
\IEEEauthorblockN{Beihong Jin\textsuperscript{\Letter} \thanks{\Letter \quad
 Corresponding author.}}
\IEEEauthorblockA{\textit{Institute of Software}\\
\textit{Chinese
Academy of Sciences}\\
\textit{ University of Chinese Academy of Sciences}\\
Beijing, China \\
Beihong@iscas.ac.cn}
\and
\IEEEauthorblockN{Yisong Yu}
\IEEEauthorblockA{\textit{Institute of Software} \\
\textit{Chinese
Academy of Sciences}\\
Beijing, China \\
yuyisong20@otcaix.iscas.ac.cn}
\and
\IEEEauthorblockN{ Yiyuan Zheng}
\IEEEauthorblockA{\textit{Institute of Software} \\
\textit{Chinese
Academy of Sciences}\\
Beijing, China \\
zhengyiyuan22@otcaix.iscas.ac.cn}
\and
\IEEEauthorblockN{ Jiageng Song}
\IEEEauthorblockA{\textit{Institute of Software} \\
\textit{Chinese
Academy of Sciences}\\
Beijing, China \\
songjiageng20@otcaix.iscas.ac.cn}
\and
\IEEEauthorblockN{ Wei Zhuo}
\IEEEauthorblockA{\textit{MX Media Co., Ltd} \\
Singapore \\
zhuoweimx@163.com}
\and
\IEEEauthorblockN{Tao Xiang}
\IEEEauthorblockA{\textit{College of Computer Science} \\
\textit{Chongqing University}\\
Chongqing, China \\
txiang@cqu.edu.cn}
}


%


\maketitle



%
\IEEEpeerreviewmaketitle

\begin{abstract}
Watching micro-videos is becoming a part of public daily life. Usually, user watching behaviors are thought to be rooted in their multiple different interests. In the paper, we propose a model named OPAL for micro-video matching, which elicits a user’s multiple heterogeneous interests by disentangling multiple soft and hard interest embeddings from user interactions. Moreover, OPAL employs a two-stage training strategy, in which the pre-train is to generate soft interests from historical interactions under the guidance of orthogonal hyper-categories of micro-videos and the fine-tune is to reinforce the degree of disentanglement among the interests and learn the temporal evolution of each interest of each user. We conduct extensive experiments on two real-world datasets. The results show that OPAL not only returns diversified micro-videos but also outperforms six state-of-the-art models in terms of recall and hit rate.
\end{abstract}

\begin{IEEEkeywords}
Recommendation, multi-interest recommendation, micro-video matching
\end{IEEEkeywords}

\section{Introduction}

In recent years, micro-video apps such as TikTok, Kwai, MX TakaTak, etc. have become increasingly popular, which makes the explosive growth in the number of micro-videos. In order to generate personalized recommendations from millions of micro-video candidates effectively, the micro-video recommendation usually includes two phases, i.e., matching and ranking, as with existing large-scale recommender systems \cite{YouTubeDNN}, where matching aims to quickly recall hundreds or thousands of micro-videos from millions or billions of candidates, and ranking aims to refine the matching results and determine the final recommendations. In this paper, we focus on micro-video matching. 

\begin{figure}[t]
    \centering
    \includegraphics[height=4cm]{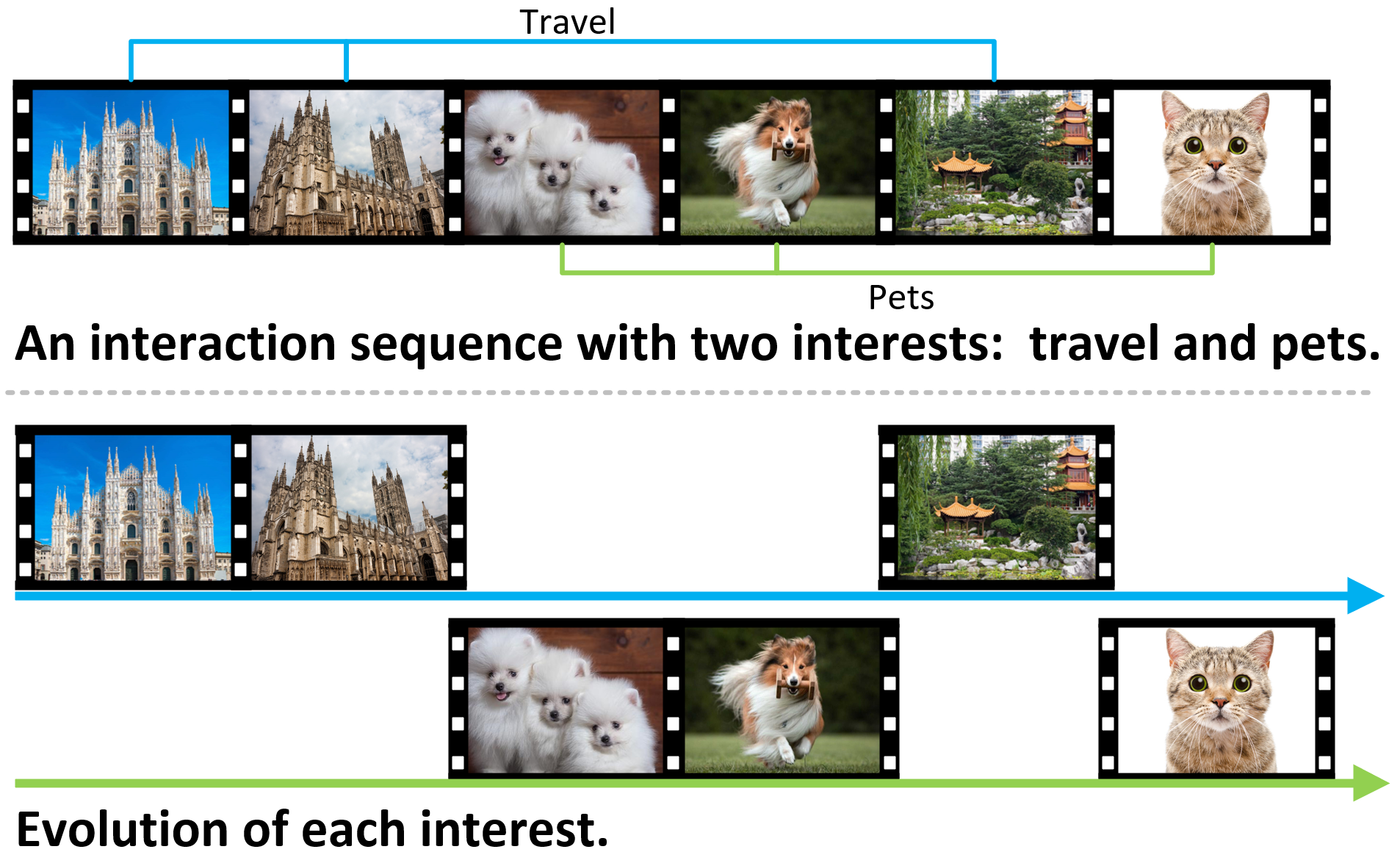}
    \caption{An example: multiple interests and interest evolution }
	\label{fig:example_fig}
\end{figure}

An ideal micro-video matching model is supposed to possess the following characteristics: (1) \textbf{Low computational complexity}. The matching process must efficiently sift through millions or billions of items within several milliseconds. Hence, the computational complexity of the matching model should be as low as possible. (2) \textbf{Recommendations with high coverage of user interests}. A user typically has diverse interests. As depicted in Figure \ref{fig:example_fig}, a user may exhibit interests in micro-videos falling under categories such as travel and cute pets. These implicit yet crucial interests influence a user's decisions regarding which micro-videos to watch. Consequently, the recalled items are supposed to encompass as many user interests as possible. (3) \textbf{Temporal dynamics of user interests}. The scope and depth of each user interest might evolve over time. For instance, a user fond of travel initially watches numerous micro-videos showcasing European scenery but gradually shifts the focus towards Asian scenery, as illustrated in Figure \ref{fig:example_fig}. Therefore, the matching model needs to accurately capture fluctuations in user interests and make predictions based on their current preferences.

There are a number of recent micro-video recommendation models available 
\cite{zeng_personalized_2018,liu_sequential_2019, wei_mmgcn_2019, liu_user-video_2019}, but most of them are appropriate for the ranking phase rather than the matching phase due to their high time complexity \cite{liu_sequential_2019,shang2023learning}. On the other hand, most of existing micro-video recommendation models offer no disentanglement of multiple interests of users \cite{chen_temporal_2018,MARS,li_routing_2019,lai2023vlogger,yu2022improving} or ignore the evolution of user interest. Though some models \cite{jiang_what_2020} learn multiple user interests, they aggregate them into one embedding while serving, limiting the contributions of multiple interests on recommendation results.

An intuitive approach to micro-video matching is to directly apply existing multi-interest recommendation models to micro-video recommendation scenarios. Existing multi-interest recommendation models utilize attention mechanisms \cite{ kang_self-attentive_2018, cen_controllable_2020}, dynamic routing \cite{ li_multi-interest_2019, cen_controllable_2020}, and contrastive learning \cite{ma_disentangled_2020,Re4,CMI} to extract multiple user interests. However, these models either fail to ensure the heterogeneity among interests or ignore the temporal dynamics of user interests, affecting the diversity and accuracy of recommendation.

In order to address the above issues and challenges, in the paper, we propose an effective and efficient model named \textbf{OPAL}(\textbf{O}rthogonal category and \textbf{P}ersonalized interest \textbf{A}ctivated \textbf{L}ever) for matching micro-videos, where effectiveness comes from disentanglement of heterogeneous multiple interests and modeling of interest evolution, and efficiency is supported by the low computational complexity. Our contributions are summaried as follows.

(1) We propose a heterogeneous multi-interest recommendation model OPAL for micro-video matching, which recognizes soft and hard user interests according to implicit learnable orthogonal hyper-categories of micro-videos.

(2) We present a novel two-stage interest learning strategy for OPAL, where OPAL improves the confidence of interest disentanglement in the first stage, while it considers the interest evolution and achieves the complete disentanglement of user interests in the second stage.

(3) We conduct extensive experiments on real-world datasets. The results show that OPAL outperforms six state-of-the-art models in terms of recall and hit-rate, demonstrating that identifying heterogeneous multiple user interests improves both recommendation accuracy and diversity.

\section{Methodology}

\subsection{Overview}

\begin{figure}[t]
	\centering
	\includegraphics[height=8.5cm]{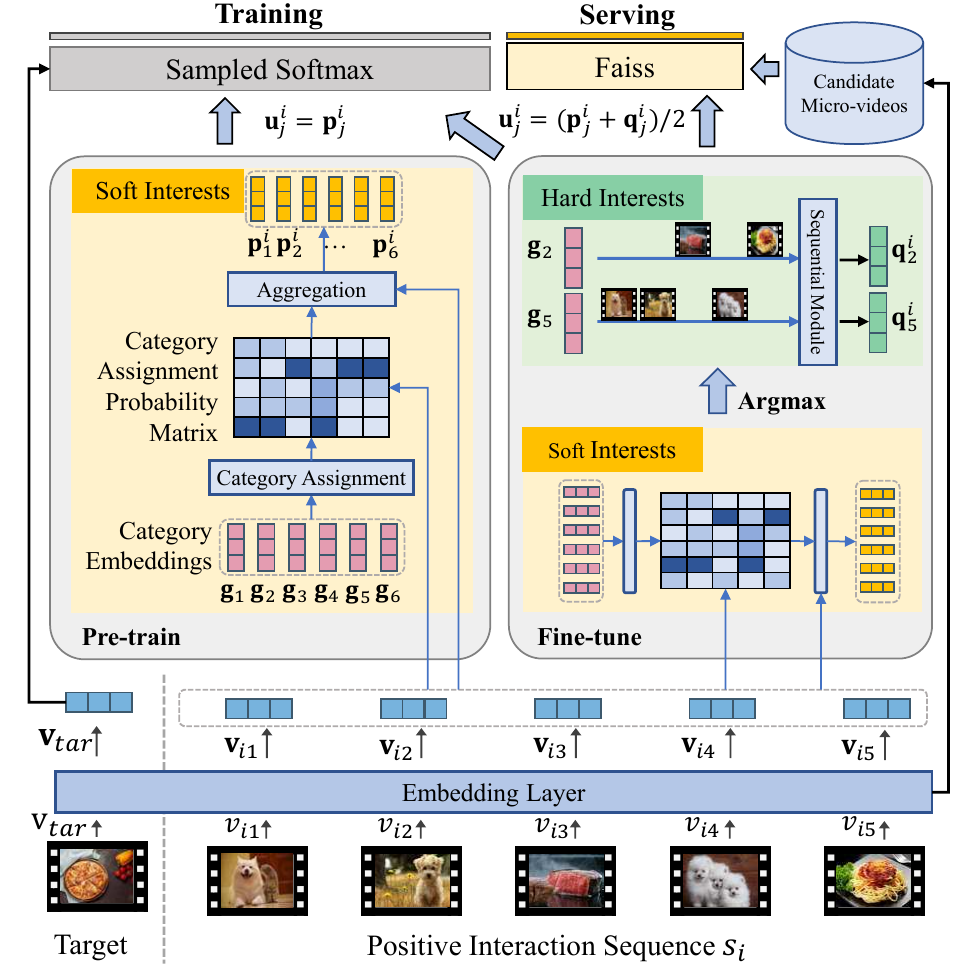}
	\caption{The architecture of OPAL, where $k=6$.}
	\label{fig:model_fig}
\end{figure}

We denote the user set as $\mathcal{U}$, the micro-video set as $\mathcal{V}$, and the embedding of a micro-video $v_i\in\mathcal{V}$ as $\mathbf{v}_i$, where $||\mathbf{v}_*||_2=1$. Figure \ref{fig:model_fig} shows the architecture of OPAL. For a user $u_i\in{\mathcal{U}}$, we first filter out micro-videos with his/her positive interactions (including interactions reflecting satisfaction such as favorites or positive engagement like replays, hereinafter referred to as interactions). Next, we sort these interactions according to the corresponding interaction timestamps to form  the interaction sequence $s_i = [v_{i1}, v_{i2},\ldots, v_{i|s_i|}]$, where $|s_i|$ denotes the length of the sequence. Then, $n$ interest $\mathbf{U}_i= [\mathbf{u}_i^1, \mathbf{u}_i^2, \ldots, \mathbf{u}_i^n] \in \mathbb{R}^{d\times{n}}$ are learned from $s_i$. Finally, we use each interest embedding to retrieve a number of candidate micro-videos via Faiss\footnote{https://github.com/facebookresearch/faiss}, and the micro-videos with the top $K$ highest matching scores are the recommendations.


The training process of OPAL includes two stages, i.e., pre-train and fine-tune, where we handle interest disentanglement via different strategies. Firstly, at the pre-train stage, each interacted micro-video is allowed to be associated with multiple interests of a user to learn multiple soft interests. We train the model until convergence to improve the confidence of interest disentanglement. Then, at the fine-tune stage, on the basis of the pre-trained parameters, the hard interests are fused to fine-tune the model until the model converges again. At this time, each micro-video is routed to one of the multiple interests exclusively to obtain the heterogeneous interests and achieve non-overlapping disentanglement of interests. Apart from that, sequential modules are adopted to excavate the evolution of each interest.

\subsection{Multi-interest Modeling}


\noindent\textbf{Global Hyper-category Embeddings.} Micro-videos can be divided into several hyper-categories, and each of them corresponds to one user interest. In order to mine semantic interests, we learn the global implicit  micro-video hyper-categories and set up $k$ micro-video hyper-category embeddings as $\{\mathbf{g}_1, \mathbf{g}_2, \ldots, \mathbf{g}_k\}$, where $\mathbf{g}_*\in{\mathbb{R}}^d$ and $||\mathbf{g}_*||_2=1$. In order to keep the heterogeneity between different hyper-categories of micro-videos and reduce the information redundancy among multiple user interests, we keep any pair of micro-video hyper-category embeddings orthogonal. Therefore, we construct the orthogonal constraint loss as follows,

\begin{equation}\label{eq:l-orth}
\mathcal{L}_{{orth}}=\sum\nolimits_{i=1}^{k} \sum\nolimits_{j=1, j \neq i}^{k} (\mathbf{g}_{i}^{T} \mathbf{g}_{j})^2.
\end{equation}


The micro-video hyper-category embeddings can be regarded as a set of unit orthogonal basis vectors. Then, in the multidimensional space spanned by orthogonal basis vectors, for a micro-video $v_i\in\mathcal{V}$, we can obtain its coordinates i.e., cosine similarities  $\left\{\mathbf{g}_{1}^{T} \mathbf{v}_{i}, \mathbf{g}_{2}^{T} \mathbf{v}_{i}, \ldots, \mathbf{g}_{k}^{T} \mathbf{v}_{i}\right\}$, where $\mathbf{g}_{k}^{T} \mathbf{v}_{i}\in[-1, 1]$. The coordinates measure the correlation between the micro-video $v_i$ and each hyper-category. The larger the coordinate, the stronger the correlation.


\noindent\textbf{Multiple Interests at the Pre-train Stage.} Given the interaction sequence of user ${u_i}$, we first calculate the coordinates of the $l$-th interacted micro-video $v_{il}\in{s_i}$ on each hyper-category $g_j$, as $r_{lj}^i=\mathbf{g}_j^T \mathbf{v}_{il}$, where $j\in\{1,2,\ldots,k\}$. Next, we convert the coordinates into probabilities for hyper-category assignment via softmax function. If we perform softmax transformation on $r_{lj}^i\in[-1,1]$ directly, the  probabilities of assigning a micro-video to the most relevant hyper-category (the coordinate value is 1) and to the most irrelevant hyper-category (the coordinate value is -1) will be very close, which smooths out the differences. Therefore, we amplify the coordinate values by dividing a small number $\epsilon$ that is less than 1. Thus, the probability of assigning $v_{il}$ to the $j$-th hyper-category is calculated as Equation \ref{eq:q_lji}.

\begin{equation}\label{eq:q_lji}
a_{l j}^{i}=\frac{\exp \left(\mathbf{g}_{j}^{T} \mathbf{v}_{i l} / \epsilon\right)}{\sum_{j=1}^{k} \exp \left(\mathbf{g}_{j}^{T} \mathbf{v}_{i l} / \epsilon\right)}
\end{equation}

Then, due to the presumption that frequent positive interactions with a hyper-category of micro-videos reflect an interest, we aggregate the embeddings of the micro-videos assigned to the same hyper-category to obtain the corresponding \textbf{soft interest}. For example, the soft interest  $\mathbf{p}_j^i$ of user $u_i$ learned according to hyper-category $j$ is calculated as follows,

\begin{equation}\label{eq:pji}
\mathbf{p}_{j}^{i}=\sum\nolimits_{l=1}^{\left|s_{i}\right|} a_{l j}^{i} \mathbf{v}_{i l}.
\end{equation}

The intensity of user interest $\mathbf{p}_j^i$ can be measured as the total assignment probabilities accumulated from history interactions, i.e., $\sum_{l=1}^{\left|s_{i}\right|} a_{l j}^{i}$. Finally, we set the $j$-th soft interest of user $u_i$ as $\mathbf{u}_j^i=\mathbf{p}_j^i$. 


In order to avoid the trivial solution that all the micro-videos concentrate on only one certain hyper-category, we restrict the number of micro-videos absorbed by each hyper-category to be approximately equal. Since the number of micro-videos cannot be trained by backpropagation, we estimate it via the total assignment probabilities collected by each hyper-category from all candidate micro-videos, i.e., the number of micro-videos assigned to hyper-category $j$ is estimated as $w_{j}=\sum_{u_i\in{\mathcal{U}}}\sum_{v \in s_i} a_{v j}^i$. The quantity distribution of micro-videos in various hyper-categories is denoted as $\mathbf{w} = [w_1, \ldots, w_k]$. Then, we constrain the discrete coefficient of $\mathbf{w}$ to be as small as possible, and construct the uniformity loss as follows,

\begin{equation}\label{eq:l_unif}
\mathcal{L}_{u n i f}=\sigma_{\mathbf{w}} / \mu_{\mathbf{w}},
\end{equation}
where $\sigma_{\mathbf{w}}$ is the standard deviation of the vector $\mathbf{w}$ and $\mu_{\mathbf{w}}$ is the average of the vector $\mathbf{w}$. In practice, it will update the embeddings of all the micro-videos when the uniformity loss is optimized, which is pretty time-consuming. In order to accelerate, we adopt the uniformity loss on micro-videos involved in the current batch to estimate the uniformity loss on all the candidate micro-videos.


\noindent\textbf{Multiple Interests at the Fine-tune Stage.} The user interests may change over time. For example, a user previously liked micro-videos about scenery in Europe, but now turns to landscape in Asia. In this case, it is necessary to model the evolution of each interest.  However, at the pre-train stage, we have not separated interaction sequences into non-overlapping subsequences related to one interest exclusively, which brings challenges to model the evolution of each interest. Therefore, based on the high confidence of hyper-category assignment, we adopt hard hyper-category assignment to achieve complete separation of the interaction sequences.

We assign each micro-video $v_{il}$ of user $u_i$ to the hyper-category $c_{i l}$ with the highest probability exclusively, i.e., $c_{i l}=\operatorname{argmax}\limits_{0<j\leq k} a_{l j}^{i}$. In order to further increase the hyper-category assignment confidence, we establish a cross-entropy based unique loss, as shown in Equation (\ref{eq:l_unique}), which forces a micro-video to be assigned to one of the video hyper-categories exclusively.

\begin{equation}\label{eq:l_unique}
\mathcal{L}_{unique}(s_{i})=\frac{1}{\left|s_{i}\right|} \sum\nolimits_{l=1}^{\left|s_{i}\right|}-\ln \frac{\max_{0<j\leq k} \left(\left\{a_{l j}^{i} \right\}\right)}{\sum\nolimits_{j=1}^{k} a_{l j}^{i}}
\end{equation}

Based on the hard hyper-category of micro-videos, the interaction sequence $s_i$ of user $u_i$ will be split into $k$ disjoint subsequences $\{s_i^j, 1<j<k\}$, where $\left|s_{i}\right| = \sum_{j=1}^{k}\left|s_{i}^{j}\right|$. 

Furthermore, in order to capture the evolution of each interest, for each non-empty subsequence $s_i^j$, we adopt the sequential module to learn the corresponding hard interest embedding $\mathbf{q}_j^i$. For simplicity, here we utilize GRU \cite{hidasi2018recurrent}.

Finally, we set $\mathbf{u}_j^i=(\mathbf{p}_j^i + \mathbf{q}_j^i)/2$, integrating the hard interests with the soft ones to fine-tune the model, which leads the user interests into complete separation and learns the interest evolution over time.

\subsection{Loss Function}

Most existing sequential recommendation models formulate the recommendation as a next-item prediction task, that is, as for a sequence $s_{i}=\left[v_{i 1}, v_{i 2}, \ldots, v_{i\left|s_{i}\right|}\right]$, they utilize $\left[v_{i 1}, v_{i 2}, \ldots, v_{i t}\right]$ to predict $v_{i(t+1)}$, where $t\in[1, |s_i|]$. In this situation, $v_{i(t+1)}$ is regarded as the positive sample, other candidate micro-videos are regarded as negative ones. However, after time $t$, the user may interact with several micro-videos $f_{i}=\left[v_{i(t+1)}, \ldots, v_{i\left(t+\left|f_{i}\right|\right)}\right]$, where $|f_{i}|\geq 1$. Regarding micro-videos other than $v_{i(t+1)}$ as negative samples is too aggressive, suppressing some micro-videos the user is interested in. Therefore, we formulate the micro-video recommendation problem as a future-item prediction task, that is, each micro-video in the future positive interaction is regarded as a positive sample.

For a user $u_i$, we randomly sample a video $v_{pos}^i\in{\left[v_{i(t+1)}, \ldots, v_{i\left(t+\left|f_{i}\right|\right)}\right]}$ as the positive sample $v_{pos}^i$ of $s_i$. Next, we randomly sample videos that have not been interacted from all the micro-video candidates to form a negative sample set $\mathcal{N}$. Then, we construct the cross-entropy loss as shown in Equation \ref{eq:l_main}.

\begin{equation}\label{eq:l_main}
\mathcal{L}_{main}= \\-\ln \frac{\exp \Big(\max\limits_{0<j < k} \left(\mathbf{v}_{\text {pos}}^{i T} \mathbf{u}_{j}^{i} / \epsilon\right)\Big)}{\sum_{v \in \mathcal{N}\cup{\{v_{pos}^i\}}} \exp \Big(\max\limits_{0<j<k} \left(\mathbf{v}^{T} \mathbf{u}_{j}^{i} / \epsilon \right)\Big)}
\end{equation}

In order to avoid the additional overhead caused by negative sampling, an intuitive operation is to treat other positive samples within the same minibatch as negative samples. But in this way, the diversity of negative samples is very limited. Since matching aims to exclude plenty of obviously irrelevant micro-videos, we need random sampling among all the micro-video pools to get richer negative micro-videos. To balance the sampling overhead and negative sample diversity, as for a batch of training samples, we sample a negative micro-video for each one. Assuming the batch size is $B$, we have $B$ negative micro-videos and share them in the batch. So, we calculate multi-class classification probabilities via the sampled softmax on $B+1$ micro-videos, where one is positive and others are negative.

In the end, the loss function is defined as Equation \ref{eq:final_loss}, where $\lambda_*$ are the regularization coefficients.

\begin{equation}\label{eq:final_loss}
L=\mathcal{L}_{main}+\lambda_{o} \mathcal{L}_{orth}+\lambda_{f} \mathcal{L}_{unif}+\lambda_{q} \mathcal{L}_{unique}
\end{equation}

\subsection{Complexity Analysis}

The parameters of OPAL are mainly the embeddings of micro-videos and hyper-categories, whose number is $|\mathcal{V}|d+kd$. The computation of OPAL focuses on interest elicitation. For an interaction sequence $s$, the time complexity of learning soft and hard interests is $O(|s|kd)$ and $O(|s|kd+|s|d^2)$, respectively. The low complexity ensures that OPAL can support large-scale micro-video matching.

\section{Experiments}
\subsection{Experiment Setup}
\noindent\textbf{Datasets} We conduct experiments on two datasets, i.e.,  \textbf{WeChat-Channels} and \textbf{MX-TakaTak}. WeChat-Channels is a dataset released by WeChat Big Data Challenge 2021\footnote{https://algo.weixin.qq.com/}. It contains two-week interactions (including  engagement interactions such as watching and satisfaction interactions such as likes and favorites) from 20,000 anonymous users on WeChat Channels,  a popular micro-video platform in China. MX-TakaTak is collected from MX TakaTak, one of the largest micro-video platforms in India. We collect real historical interaction records of TakaTak from Sept. 18, 2021 to Sept. 28, 2021, then randomly sample 50,000 users and form the MX-TakaTak dataset using their interactions. The statistics of the two datasets are shown in Table \ref{tab:dataset-stat}.

To avoid the data penetration during the model training and evaluation \cite{wu2019session}, we strictly maintain the timing relationship among the two datasets. In detail, we construct the test set using interactions of the last day, the validation set using interactions of the second-to-last day, and the training set using other interactions.

\noindent\textbf{Metrics} Recall@K and HitRate@K are used as metrics to evaluate the quality of the recommendation results. The value of K is set to 50 or 100 or 200.

\noindent\textbf{Competitors} We choose the following six  multi-interest  recommendation models proposed in recent years as competitors: 1) Dynamic routing-based models: \textbf{MIND} \cite{li_multi-interest_2019} and \textbf{ComiRec-DR} \cite{cen_controllable_2020}; 2) Attention-based models: \textbf{ComiRec-SA} \cite{cen_controllable_2020} and \textbf{Octopus} \cite{liu_octopus_2020}; 3) Contrastive learning-based models: \textbf{DSSRec} \cite{ma_disentangled_2020} and \textbf{Re4} \cite{Re4}.

\noindent\textbf{Implementation Details} We implement OPAL with PyTorch, initialize the parameters with the uniform distribution $U(-\frac{1}{\sqrt{d}}, \frac{1}{\sqrt{d}})$, and optimize the model through Adam. The learning rate is searched in $[0.1, 0.01, 0.001, 0.0001]$, and finally set to 0.001, $\lambda_{q}$, $\lambda_{o}$, $\lambda_{f}$ are searched in $[0.5, 1, 5, 10, 20]$, and finally set $\lambda_{q}=1$, $\lambda_{o}=10$, $\lambda_{f}=1$, and the hyperparameter $\tau$ is set to 0.1. Our code is available at https://github.com/libeibei95/OPAL.

For the sake of fairness, we set the embedding dimension in all the experiments to 64 and batch size to 1024. We stop training when Recall@200 on the validation set has not been improved in 5 consecutive epochs. As for multi-interest based models, we search for the number of interests in $[1, 2, 4, 8, 16]$, and select the one with best performance on Recall@200. Note that we train all the competitors to optimize the future-item prediction task. Besides, for MIND, ComiRec-SA, ComiRec-DR and DSSRec, we use the open-source code released on Github\footnote{https://github.com/Wang-Yu-Qing/MIND}$^,$\footnote{https://github.com/THUDM/ComiRec}$^,$\footnote{https://github.com/abinashsinha330/DSSRec}.

\begin{table}[t]
\scriptsize
\begin{center}
\begin{tabular}{l|c|c|c|r}
  \hline
  Dataset & \#Users & \#Micro-videos & \#Interactions & Density  \\
  \hline
  WeChat-Channels & 20000 & 77557 & 2666296 & 0.17\% \\
  MX-TakaTak & 50000 &	157691	& 33863980	& 0.45\% \\
  \hline
\end{tabular}
\end{center}
\caption{The statistics of the two datasets.} \label{tab:dataset-stat}
\end{table}

\begin{table*}[tb]
\scriptsize
\begin{center}
\begin{tabular}{l|c|c c c | c c c|c|c c c | c c c}
\hline
&  \multicolumn{7}{|c|}{WeChat-Channels}& \multicolumn{7}{|c}{MX-TakaTak}\\
\cline{2-15}
&  & \multicolumn{3}{|c|}{Recall}& \multicolumn{3}{|c|}{HitRate}&  & \multicolumn{3}{|c|}{Recall}& \multicolumn{3}{|c}{HitRate}\\
\hline
 & \textbf{\#I.} & \textbf{@50} & \textbf{@100} & \textbf{@200} & \textbf{@50} & \textbf{@100} & \textbf{@200} & \textbf{\#I.} & \textbf{@50} & \textbf{@100} & \textbf{@200} & \textbf{@50} & \textbf{@100} & \textbf{@200}\\
\hline
\small{Octopus} & 1 & 0.0597 & 0.1111 & 0.1852 & 0.3382 & 0.4953 & 0.6367 & 1 & 0.0403 & 0.0863 & 0.1549 & 0.3872 & 0.5519 & 0.6756 \\
\small{MIND} & 1 & 0.1149 & 0.1822 & 0.2774 & 0.5073 & 0.6370 & 0.7524 & 8 & 0.0793 & 0.1246 & 0.1847 & 0.5071 & 0.6434 & 0.7442\\
\small{ComiRec-DR} & 1 & 0.1149 & 0.1831 & 0.2744 & 0.5089 & 0.6411 & 0.7508 & 1 & 0.0758 & 0.1203 & 0.1824 & 0.5299 & 0.6533 & 0.7395\\
\small{ComiRec-SA} & 1 & \underline{0.1215} & \underline{0.1934} & \underline{0.2887} & 0.5228 & \underline{0.6603} & 0.7633 & 1 & 0.0740 & 0.1190 & 0.1794 &\underline{0.5402} & \underline{0.6567} & \underline{0.7412}\\
\small{DSSRec} & 8 & 0.1200 & 0.1879 & 0.2836 & \underline{0.5274} & 0.6539 & \underline{0.7640} & 1 & \underline{0.0844} & \underline{0.1261} & \underline{0.1882} & 0.5253 & 0.6345 & 0.7288\\
\small{Re4} & 1 & 0.0948 & 0.1548 & 0.2386 & 0.4415 & 0.5770 & 0.6993 & 4 & 0.0749 & 0.1171 & 0.1769 & 0.5300 & 0.6346 & 0.7263\\
\hline
\small{OPAL} & 4 & \textbf{0.1424} &\textbf{0.2183} & \textbf{0.3140} & \textbf{0.5744} & \textbf{0.6973} & \textbf{0.7881} & 8 & \textbf{0.0903} & \textbf{0.1373} & \textbf{0.2021} & \textbf{0.5653} & \textbf{0.6736} & \textbf{0.7598}\\
\hline
\small{Improv.(\%)}&  & 17.23  & 12.87  & 8.77  & 8.91  & 5.61  & 3.15  &  & 6.99  & 8.88  & 7.39  & 4.65  & 2.58  & 2.51 \\
\hline
\end{tabular}
\end{center}
\caption{Recommendation accuracy on two datasets. \#I. denotes the number of interests. The number in a bold type is the best performance in each column. The underlined number is the second best in each column.} \label{tab:exp}
\end{table*}

\subsection{Performance Comparison}

We conduct a comparative experiment, comparing our model with six competitors. Setting various numbers of interests, we record the best performance of each model in Table \ref{tab:exp}. From Table \ref{tab:exp}, we have the following observations. (1) OPAL surpasses the competitors in all metrics on the two datasets. For example, on the WeChat-Channel dataset, OPAL outperforms the second best model, i.e., ComiRec-SA, about 17.2\% on Recall@50, and 8.77\% on HitRate@50. (2) Although the competitors claim they are multi-interest models, most of them have the best recommendation performance while learning only one interest for per user. This finding shows that these models cannot mine meaningful multiple user interests in micro-video scenarios. (3) The Octopus model is very similar to the hard interest part of OPAL, but its performance is far from the performance of OPAL.

\begin{table}[t]
\scriptsize
\begin{center}
\setlength\tabcolsep{3pt}
\begin{tabular}{l|l|c c c | c c c}
\hline
 & & \multicolumn{3}{|c|}{Recall}& \multicolumn{3}{|c}{HitRate}\\
\cline{2-8}
 Datasets & {Models} & \textbf{@50} & \textbf{@100} & \textbf{@200} & \textbf{@50} & \textbf{@100} & \textbf{@200} \\
 \hline
 \multirow{7}{*}{WeChat}      & (A) w/o $\mathcal{L}_{unique}$ & 0.1281 & 0.2001 & 0.2937 & 0.5131 & 0.6464 & 0.7544\\
        & (B) w/o $\mathcal{L}_{orth} $ & 0.1216 & 0.1890 & 0.2859 & 0.5022 & 0.6302 & 0.7459\\
        & (C) w/o $\mathcal{L}_{unif}$ & 0.1272 & 0.1965 & 0.2865 & 0.5156 & 0.6433 & 0.7508\\
        \cline{2-8}
        & (D) w/o pre-train & 0.0794 & 0.1370 & 0.2238 & 0.4141 & 0.5579 & 0.6901\\
        & (E) w/o fine-tune & 0.1234 & 0.1901 & 0.2810 & 0.5369 & 0.6580 & 0.7611\\
        \cline{2-8}
        & (F) next-item  & 0.1204 & 0.1828 & 0.2696 & 0.5245 & 0.6487 & 0.7516\\
        \cline{2-8}
        & (G) OPAL & \textbf{0.1424} & \textbf{0.2183} & \textbf{0.3140} & \textbf{0.5744} & \textbf{0.6973} & \textbf{0.7881}\\
        \hline
 \multirow{7}{*}{TakaTak}
 & (A) w/o $\mathcal{L}_{unique}$ & 0.0855 & 	0.1314 & 	0.1947 & 	0.5381 & 	0.6487 & 	0.7407\\
  & (B) w/o $\mathcal{L}_{orth} $ & 0.0886 & 	0.1337 & 	0.1988 & 	0.5579 & 	0.6658 & 	0.7548\\
  & (C) w/o $\mathcal{L}_{unif}$ & 0.0820 & 	0.1217 & 	0.1933 & 	0.5119 & 	0.6332 & 	0.7323\\

\cline{2-8}
& (D) w/o pre-train  & 0.0713 & 0.1177 & 0.1810 & 0.5164 & 0.6254 & 0.7213\\
& (E) w/o fine-tune & 0.0794 & 0.1263 & 0.1898 & 0.4850 & 0.6121 & 0.7170\\
\cline{2-8}
& (F) next-item &  0.0664 & 0.1075 & 0.1679 & 0.4863 & 0.6034 & 0.7039\\
\cline{2-8}
& (G) OPAL & \textbf{0.0903} & \textbf{0.1373} & \textbf{0.2021} & \textbf{0.5653} & \textbf{0.6736} & \textbf{0.7598}\\
\hline
\end{tabular}
\end{center}
\caption{Ablation study.} \label{tab:ablation-study}
\end{table}

\subsection{Ablation Study}

We conduct ablation studies to investigate the effect of the three regularization losses, i.e., $\mathcal{L}_{orth}$, $\mathcal{L}_{unif}$, $\mathcal{L}_{unique}$, two-stage training strategy and the future-item prediction task. The experimental results are shown in Table \ref{tab:ablation-study}, from which we have the following observations and conclusions.

\noindent\textbf{Effect of Regularization Losses.} From model variants (A)-(C), we find that whichever loss is removed leads to a decrease in the performance of the model, which indicates their contribution to the performance. This is due to the fact that all the three losses are constrained to help enhance the heterogeneity and representation quality of multiple user interests.

\noindent\textbf{Effect of Two-stage Training.} We remove the pre-train stage and the fine-tune stage from our model, respectively, denoted as model variants (D) and (E). We find that the performance gets worse after removing any of two stages. This demonstrates that the hard classification method and time series information at the fine-tune stage contribute to the performance improvement. It also implies that if we omit the pre-train stage and directly use the rigorous hard classification method, the model may converge to a local optimal solution.

\noindent\textbf{Effect of the Future-item Prediction Task.} We construct a variant of OPAL that replaces the future-item prediction task with the next-item prediction task, denoted as model variant (F). We find that the intact OPAL with the future-item prediction task is more effective than the variant with the next-item prediction task. This is because that predicting the user interactions over the next period of time is more close to the requirement of real-world scenarios.

\begin{figure}[tb]
	\centering
	\includegraphics[width=0.45\textwidth]{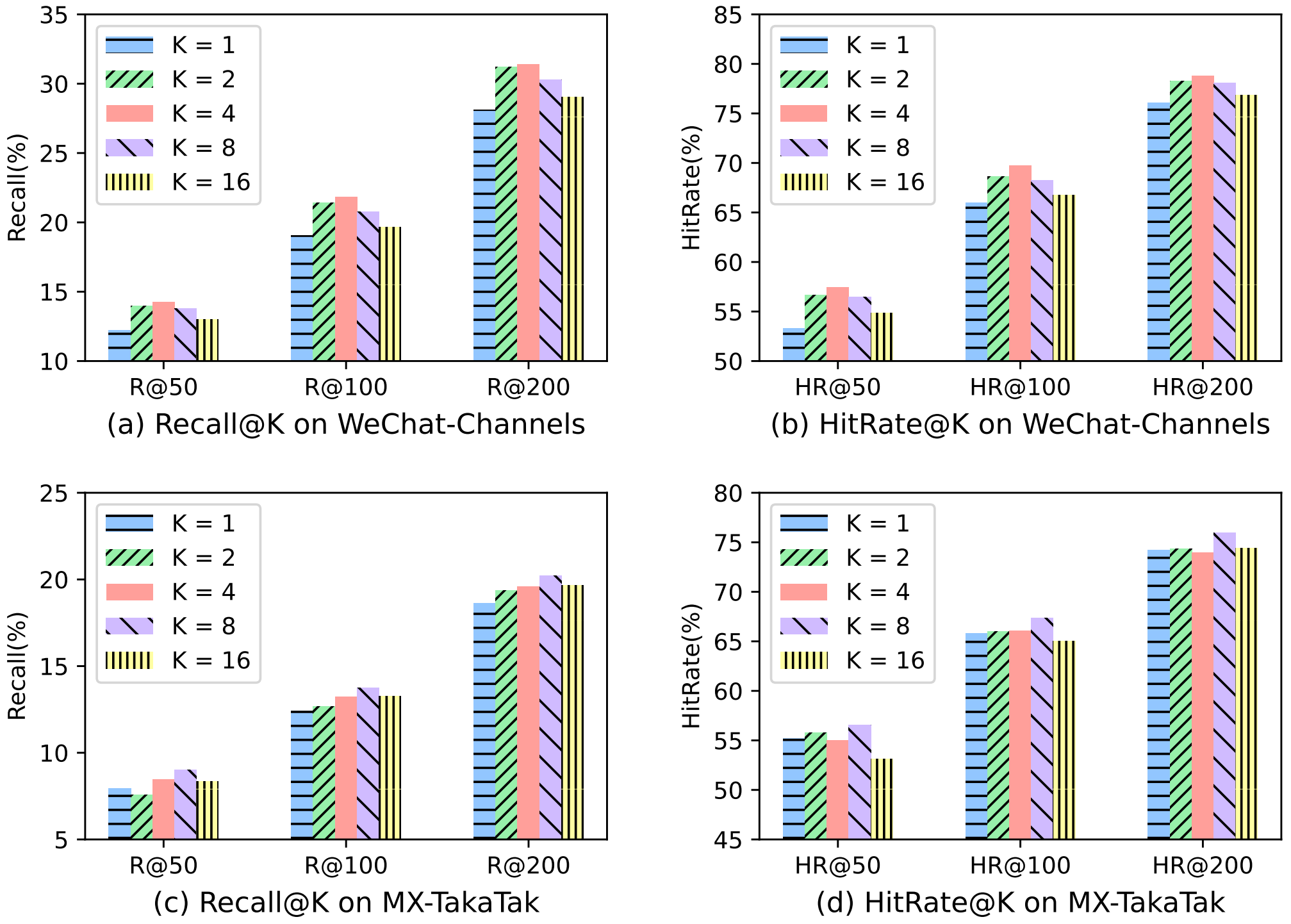}
	\caption{The impact of multiple interests on performance}
	\label{fig:ninterests}
\end{figure}

\subsection{Multiple Interests Analysis}

\noindent\textbf{Impact on Recommendation Accuracy.} In order to study the impact of the number of interests in our model on the recommendation accuracy, we set the number of micro-video hyper-categories to  1, 2, 4, 8, 16 successively and conduct experiments. The experimental results are shown in Figure \ref{fig:ninterests}. It can be seen that our model on the WeChat-Channel dataset reaches the best when the number of interests is 4, and reaches the best on the MX-TakaTak dataset when the number of interests is 8. This shows that our model is reasonable and effective to disentangle multiple interests from user interaction sequences and utilize them to achieve recommendations.

\begin{figure}[tb]
	\centering
	\includegraphics[width=0.45\textwidth]{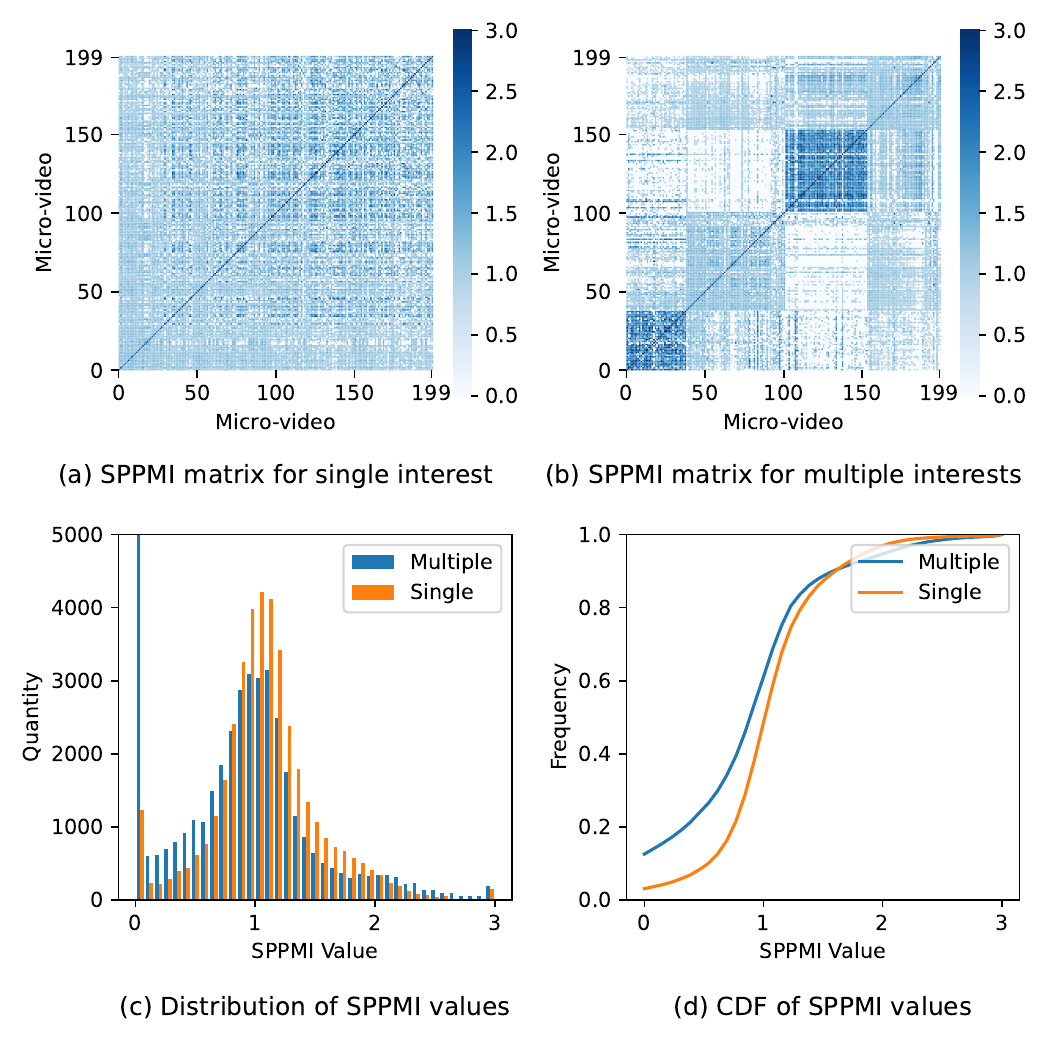}
	\caption{The impact of multiple interests on diversity}
	\label{fig:visual-diversity}
\end{figure}

\noindent\textbf{Impact on Recommendation Diversity.} We conduct a case study to compare the recommendation diversity of single-interest modeling with multi-interest modeling. We randomly sample a user. After setting the number of interests to 1 and 4, OPAL outputs 200 micro-videos as recommendation results, respectively. In particular, when the number of interests is 4, the number of micro-videos recalled by each interest embedding is 38, 63, 52 and 47, respectively. Since the labels of micro-videos are often missing or inaccurate, we use SPPMI value \cite{levy2014neural} to measure the correlation between micro-videos. The higher the SPPMI value, the greater the correlation. We calculate SPPMI values between two micro-videos in the recommendation results and get a SPPMI matrix with dimensions 200$\times$200. The visualization results are shown in Figures \ref{fig:visual-diversity} (a)-(b). We can see that the micro-videos recommended by the single interest model are  highly correlated. In the recommendation result of the multi-interest model, the micro-videos recalled by the same interest embedding have high correlations, while the micro-videos recalled by different interests have low correlations. These demonstrate that the heterogeneity of recommended micro-videos is brought by different interest embeddings. Moreover, from Figures \ref{fig:visual-diversity}(c)-(d), we can find that there are more small values in the multi-interest SPPMI matrix compared with the single-interest SPPMI matrix, which indicates that the recommendation results from the multi-interest model have greater diversity.

\section{Conclusion}

This paper proposes a micro-video recommendation model OPAL. OPAL relys on hyper-categories information of micro-videos in positive interaction sequences to distinguish multiple interests for the user and adopts pre-train plus fine-tune to form multiple soft and hard interest embeddings of the user. In particular, in OPAL, the recommendation results are recalled using Faiss with multiple interest embeddings as input, increasing the diversity of micro-videos, and the future-item prediction task, which is closer to the real scenarios, is introduced to optimize the model, improving the recommendation performance. Results of experiments on real-world datasets demonstrate the feasibility and effectiveness of OPAL.

\section*{Acknowledgment}

This work was supported by the National Natural Science Foundation of China
under Grant No. 62072450.

\end{document}